\documentclass[%
reprint,
showpacs,
bibnotes,
amsmath,amssymb,
aps,
prb,
floatfix,
]{revtex4-1}

\usepackage{graphicx}
\usepackage{dcolumn}
\usepackage{bm}
\usepackage{hyperref}
\usepackage{multirow}
\usepackage{array}
\usepackage{booktabs}
\usepackage{ctable}
\usepackage{upgreek}
\usepackage{epsfig}
\usepackage{mathrsfs}
\usepackage{amssymb}
\usepackage{amsbsy}
\usepackage{color}
\usepackage{cancel}
\usepackage{pifont}
\usepackage{marginnote}
\newcommand{\bcen}{\begin{center}}
\newcommand{\ecen}{\end{center}}
\newcommand{\btab}{\begin{tabular}}
\newcommand{\etab}{\end{tabular}}
\newcommand{\bdes}{\begin{description}}
\newcommand{\edes}{\end{description}}

\newcommand{\beq}{\begin{equation}}
\newcommand{\eeq}{\end{equation}}
\newcommand{\bea}{\begin{eqnarray}}
\newcommand{\eea}{\end{eqnarray}}

\newcommand{\half}{\frac{1}{2}}
\newcommand{\bary}{\begin{array}}
\newcommand{\eary}{\end{array}}
\newcommand{\benum}{\begin{enumerate}}
\newcommand{\eenum}{\end{enumerate}}
\newcommand{\bitem}{\begin{itemize}}
\newcommand{\eitem}{\end{itemize}}

\newcommand{\blam}{{\boldsymbol{\lambda}}}

\newcommand{\bDelta}{{\boldsymbol{\Delta}}}
\newcommand{\bPi}{{\boldsymbol{\Pi}}}

\newcommand{\bGam}{{\boldsymbol{\Gamma}}}

\newcommand{\be} { \mbox{\boldmath $e$}}

\newcommand{\bk} { \bm{k} }

\newcommand{\bq} { \bm{q} }

\newcommand{\dou}{\partial}

\newcommand{\eqn}[1] {eqn.~(\ref{#1})}

\newcommand{\fig}[1]{fig.~\ref{#1}}
\newcommand{\Fig}[1]{Fig.~\ref{#1}}

\makeatletter

\newcommand{\Rmnum}[1]{\expandafter\@slowromancap\romannumeral #1@}
\makeatother

\newcommand{\myfigwidth}{0.4\paperwidth}

\newcommand{\as}{a_{s}}

\newcommand{\kf}{k_F}

\newcommand{\mylabel}[1]{\label{#1}} 

\newcommand{\myonlinecite}[1]{[\onlinecite{#1}]}
\newcommand{\mycite}[1]{\cite{#1}}

\usepackage{lineno}

\begin{document}



\title{Collective excitations across the BCS-BEC crossover induced by a synthetic Rashba spin-orbit coupling}

\author{Jayantha P.~Vyasanakere}
\email{jayantha@physics.iisc.ernet.in}
\author{Vijay B.~Shenoy}
\email{shenoy@physics.iisc.ernet.in}
\affiliation{Centre for Condensed Matter Theory, Department of Physics, Indian Institute of Science, Bangalore 560 012, India}



\date{\today}

\begin{abstract}
Synthetic non-Abelian gauge fields in cold atom systems produce a
generalized Rashba spin-orbit interaction described by a vector $\blam
= (\lambda_x, \lambda_y, \lambda_z)$ that influences the motion of
spin-$\half$ fermions. It was recently shown [Phys.~Rev.~B 84, 014512
  (2011)] that on increasing the strength of the spin-orbit coupling
$\lambda = |\blam|$, a system of fermions at a finite density
$\rho\approx\kf^3$ evolves to a BEC like state even in the presence of
a weak attractive interaction (described by a scattering length
$\as$). The BEC obtained at large spin-orbit coupling ($\lambda \gg
k_F$) is a condensate of rashbons -- novel bosonic bound pairs of
fermions whose properties are determined solely by the gauge field. In
this paper, we investigate the collective excitations of such
superfluids by constructing a Gaussian theory using functional
integral methods. We derive explicit expressions for superfluid phase
stiffness, sound speed and mass of the Anderson-Higgs boson that are
valid for any $\blam$ and scattering length. We find that at finite
$\lambda$, the phase stiffness is always lower than that set by the
density of particles, consistent with earlier work[arXiv:1110.3565]
which attributed this to the lack of Galilean invariance of the system
at finite $\lambda$. We show that there is an {\em emergent Galilean
  invariance} at large $\lambda$, and the phase stiffness is
determined by the rashbon density and mass, consistent with Leggett's
theorem. We further demonstrate that the rashbon BEC state is a
superfluid of anisotropic rashbons interacting via a contact
interaction characterized by a rashbon-rashbon scattering length
$a_R$. We show that $a_R$ goes as $\lambda^{-1}$ and is essentially
{\em independent} of the scattering length between the fermions as
long as it is nonzero. Analytical results are presented for a rashbon
BEC obtained in a spherical gauge field with $\lambda_x = \lambda_y =
\lambda_z = \frac{\lambda}{\sqrt{3}}$.

\end{abstract}


\maketitle

\section{Introduction}
\mylabel{sec:Intro}

The simulation of quantum condensed matter
systems\cite{Ketterle2008,Bloch2008,Giorgini2008} with cold atoms has
captivated the imagination and efforts of many. Some of the most
recent new developments include the
generation\mycite{Jaksch2003,Osterloh2005,Ruseckas2005,Gerbier2010,Dalibard2011}
of synthetic gauge fields in bosons\mycite{Lin2009A, Lin2009B,
  Lin2011} and realization of fermionic degeneracy in their
presence.\mycite{HuiZhaiPrivateCommunication2012}

Uniform non-Abelian gauge fields produce  spin-orbit
interactions. The physics of bosons in spin-orbit coupled system has
been investigated by many authors.\cite{Stanescu2008,Wang2010,Ho2010}
The rich physics hidden in the fermion problem was revealed by the
solution of the two-body problem given in
ref.~\myonlinecite{Vyasanakere2011TwoBody}, where it was shown that for
certain high symmetry gauge fields, a bound state appears  {\em
  even for an infinitesimal} attraction in the singlet channel.  The
key outcome of this is that a BCS-BEC crossover is induced by increasing the strength of the  gauge
field even with a weak attractive
interaction.\cite{Vyasanakere2011BCSBEC} The BEC that is realized was
shown to be a condensate of a new type of boson -- the rashbon --
whose properties are determined solely by the gauge field and not by
the scattering length characterizing the interaction between the
fermions.  This BEC realized at large gauge coupling is called the
rashbon-BEC (RBEC). Concurrently, anisotropic superfluidity of
rashbons\mycite{Hu2011}, zero-temperature BCS-BEC crossover in the
presence of Zeeman fields\mycite{Gong2011,Iskin2011a} (imbalance) was
studied, and transition temperatures were
estimated\mycite{Yu2011,Vyasanakere2011Rashbon}. Dresselhaus like
spin-orbit interaction\mycite{Han2011,Takei2011} has also been examined. Non-Abelian gauge
fields in lower dimensions and lattices have also been
investigated.\mycite{Goldman2009,He2011,Chen2012} A review of these
fast paced recent developments may be found in ref.~\myonlinecite{Zhai2011}.
Several aspects of the physics of spin-orbit coupled fermions were
reported earlier\mycite{Chaplik2006,Cappelluti2007} and were
independently discovered in the cold atoms
context.\mycite{Vyasanakere2011TwoBody,Vyasanakere2011BCSBEC}

The motivating questions for this work pertain to the properties of
the RBEC that is obtained at large gauge coupling at a fixed
scattering length $\as$. In the usual BCS-BEC crossover\mycite{Eagles1969,Leggett1980,Leggett2006,Nozieres1985,Randeria1995} in the absence
of spin-orbit interaction, the BEC state for small positive scattering
length $\as$ is a condensate of bosons (fermionic dimer
molecules). This BEC state can be described by the Bogoliubov theory of
interacting bosons\mycite{Abrikosov1965}, where the boson mass is twice the fermion
mass  and the effective boson-boson scattering length 
is proportional to $\as$.\mycite{Randeria1995,Pethick2004} Does a
similar description hold for the RBEC obtained by tuning the magnitude
of the gauge coupling?  How does rashbon-rashbon scattering enter the
description, i.~e., what is the effective rashbon-rashbon scattering
length?  

That collective excitations have interesting and unusual features was
pointed out in ref.~\myonlinecite{Zhou2011} which studied phase
stiffness $K^s$ (superfluid density) for an extreme-oblate gauge field
(see below for a definition). In the regime $\lambda \lesssim \kf$,
the $K^s$ decreases with increasing gauge coupling. However, for
$\lambda \gtrsim \kf$, $K^s$ increases and saturates as $\lambda/\kf$
attains large values. For all $\lambda$, $K^s$ is {\em less} than
$\rho/4m$, the value of phase stiffness for a superfluid without the
spin-orbit interaction, where $\rho$ the density and $m$ is the mass
of the fermions. This is attributed\mycite{Zhou2011} to the lack
of Galilean invariance in systems with synthetic non-Abelian gauge
fields (see also, ref.~\myonlinecite{Williams2012}). While this is
true, we conjecture that Galilean will be approximately restored in
the system for $\lambda \gg \kf$ when an attractive interaction, however
weak, is present. The basis of this conjecture stems from the fact
that at large $\lambda$ the system with even a weak attraction can be
thought of as a collection of rashbons which disperse
quadratically\mycite{Vyasanakere2011Rashbon}, $\varepsilon_R(\bq) = -E^R + \sum_i
\frac{q_i^2}{2 m^R_i}$, albeit with an anisotropic dispersion defined by
the direction dependent rashbon mass $m^R_i$ and $E^R$ is the rashbon binding energy, a result that is valid
for $|\bq| \ll \lambda$. This dispersion is Galilean invariant, and
therefore we expect to obtain a phase stiffness tensor $K^s_{ij} =
\frac{\rho_R}{m^R_{i}} \delta_{ij}$ (no sum on $i$), where $\rho_R= \rho/2$ is the rashbon density,  consistent with
Leggett's result\mycite{Leggett1998,Leggett2006}. Testing this
conjecture regarding {\em emergent Galilean invariance} and  answering the questions raised in the previous paragraph are the aims of this paper.

To this end, we investigate the collective excitations of superfluids
induced by non-Abelian gauge fields using a Gaussian fluctuations
theory with a functional integral framework. Our main result is that
the rashbon BEC can be described as a collection of weakly
interacting rashbons. We obtain an effective rashbon-rashbon
scattering length which we show is generically proportional to
$\lambda^{-1}$, and is {\em independent of the scattering length
  between the fermions} to leading order. In addition, we show that
the phase stiffness has precisely the form as conjectured above. The
RBEC state is a remarkable state where the effective interaction
between the emergent bosons (rashbons) is determined by the {\em
  kinetic energy} (spin-orbit coupling $\lambda$) of the constituent
fermions, and {\em not} the attraction between the fermions as long as
it is non-vanishing. Our theory also provides the phase stiffness,
speed of sound and the mass of the Anderson-Higgs boson for any gauge
coupling. 

Sec.~\ref{Formulation} outlines the functional integral framework used
in the analysis of the collective excitations and obtains general
formulae for the phase stiffness, sound speed and Anderson-Higgs mass
for a generic Rashba like spin-orbit coupled system. Results for a
spherical gauge field are discussed in sec.~\ref{SphericalResults},
and sec.~\ref{RBECdiscussion} contains a discussion of the properties
of rashbon BECs. The paper is summarized in sec.~\ref{Summary}.

\section{Formulation}
\mylabel{Formulation}

We follow closely the notation and terminology introduced in
\mycite{Vyasanakere2011TwoBody,Vyasanakere2011BCSBEC}. The Hamiltonian
of the system of interest is made up of two pieces
\beq
{\cal H} = {\cal H}_R + {\cal H}_\upsilon.
\eeq
The kinetic energy of the spin-$\half$ fermions is
\beq
{\cal H}_R = \sum_{\bk} \varepsilon_\alpha(\bk) C^\dagger_{\bk \alpha} C_{\bk \alpha}
\eeq
where, $C$s and $C^\dagger$s are fermion operators,
\beq\mylabel{eqn:epsk}
\varepsilon_\alpha(\bk) = \frac{k^2}{2} - \alpha |\bk_\lambda|,
\eeq
$\alpha = \pm 1$ is the helicity, $\bk_\lambda = \lambda_x k_x \be_x +
\lambda_y k_y \be_y + \lambda_z k_z \be_z$. The ``vector'' $\blam
\equiv (\lambda_x,\lambda_y,\lambda_z) \equiv \lambda \hat{\blam}$
describes the configuration of the gauge field that induces a
generalized Rashba spin-orbit interaction, where $\lambda = |\blam|$
is the magnitude of the gauge coupling and $\hat{\blam}$ is a unit
vector. High symmetry gauge field configurations of interest include
the extreme oblate (EO) gauge field with $\blam = \lambda
\left(\frac{1}{\sqrt{2}},\frac{1}{\sqrt{2}},0 \right)$ and the
spherical (S) gauge field which has $\blam = \lambda
\left(\frac{1}{\sqrt{3}},\frac{1}{\sqrt{3}}, \frac{1}{\sqrt{3}}
\right)$. We use units where the fermion mass $m$ and $\hbar$
are unity. We consider a finite density of fermions
$\rho$ which defines a momentum scale $\kf$ such that $\rho =
\frac{\kf^3}{3 \pi^2}$, and an energy scale $E_F=\frac{\kf^2}{2}$.

The interaction piece ${\cal H}_\upsilon$ describes an attraction in the singlet channel as
\beq
{\cal H}_\upsilon = \frac{\upsilon}{\Omega}\sum_{\bq,\bk,\bk'} C^\dagger_{(\frac{\bq}{2} + \bk) \uparrow}  C^\dagger_{(-\frac{\bq}{2} + \bk') \downarrow}  C_{\bk' \downarrow} C_{\bk \uparrow} 
\eeq
where $\Omega$ is the volume of the system, $\upsilon$ is the bare interaction parameter. The theory requires an ultraviolet cutoff $\Lambda$ which can be eliminated by using $\frac{1}{4 \pi \as} = \frac{1}{\upsilon} + \Lambda$. Using mean-field theory, it was shown in ref.~\myonlinecite{Vyasanakere2011BCSBEC} that increasing $\lambda$  induces a BCS to BEC crossover even for a  weak attractive interaction ($|\kf \as| \ll 1, \as <0$). We aim to study the collective excitations of such superfluids across this crossover.

To this end we use a functional integral framework which has been extensively used in the study of BCS-BEC crossover.\mycite{SaDeMelo1993,Randeria1995,Engelbrecht1997,Palo1999,Dupuis2004,Diener2008}
Denoting inverse temperature as $\upbeta$ and chemical potential as $\mu$, we write the action
\begin{equation} \mylabel{eqn:FullAction}
{\cal S}[\Psi] = \sum_k\Psi^\star(k) (-G_0^{-1}(k,k')) \Psi(k') + \frac{\upsilon}{\upbeta \Omega} \sum_q S^\star(q) S(q) 
\end{equation}
where 
\beq
\Psi(k) = \left( 
\begin{array}{c} 
c_+(k) \\
c^\star_+(-k) \\
c_-(k) \\
c^\star_-(-k)
\end{array} \right)
\eeq
is a Nambu vector consisting of Grassmann variables describing the fermions, $k = (ik_n, \bk)$ where $ik_n$ is a fermionic Matsubara frequency, 
{\scriptsize
\begin{equation}
G_0^{-1}(k,k') =  \left(\begin{array}{cccc}
ik_n - \xi_{+}(\bk) & 0 & 0 &0 \\
0 &   ik_n + \xi_{+}(\bk) & 0 & 0 \\
0 & 0 &  ik_n - \xi_{-}(\bk) & 0 \\
0 & 0 & 0 &  ik_n + \xi_{-}(\bk) 
\end{array}
\right) \delta_{k,k'},
\end{equation}
}
$\xi_\alpha(\bk) = \varepsilon_\alpha(\bk) - \mu$, and 
\beq
S^\star(q) = \sum_{k,\alpha \beta} A_{\alpha \beta}(\bq,\bk) c^\star_{\alpha}(\frac{q}{2} + k)c^\star_{\beta}(\frac{q}{2} - k)
\eeq
is the Fourier transform of the singlet density with $q = (iq_\ell, \bq)$, $iq_\ell$ is a bosonic Matsubara frequency. $A_{\alpha\beta}(\bq,\bk)$ is the singlet amplitude in a two particle state of $\alpha$ and $\beta$ helicities, with  centre of mass momentum $\bq$ and relative momentum $\bk$. It must be noted that $A_{\alpha\beta}(\bq,\bk)$ satisfy many symmetry properties which are used extensively in the work that follows. Moreover, care must be exercised in the definition of $A_{\alpha\beta}(\bq,\bk)$ due to the non-zero Chern flux originating from the origin of the momentum space (see ref.~\myonlinecite{Ghosh2011}).

We now introduce a Hubbard-Stratanovich pair field $\Delta(q)$ to decouple the interaction term to obtain
\beq
{\cal S}[\Psi,\Delta] = \sum_{k,k'}\Psi^*(k) (-G^{-1}(k,k')) \Psi(k') - \frac{1}{\upsilon} \sum_q \Delta^*(q) \Delta(q)
\eeq
where $G^{-1}(k,k')$ is
\beq
G^{-1}(k,k') = G_0(k,k') - \bDelta(k,k'),
\eeq 
{\scriptsize
\beq
\bDelta(k,k') = \left(
\begin{array}{cccc}
 0 & \Delta_{++}(k,k') & 0 & \Delta_{+-}(k,k') \\
 \tilde{\Delta}_{++}(k,k') & 0 &  \tilde{\Delta}_{+-}(k,k') & 0 \\
0 & \Delta_{-+}(k,k') &  0 & \Delta_{--}(k,k') \\
\tilde{\Delta}_{-+}(k,k') & 0 & \tilde{\Delta}_{--}(k,k') & 0  
\end{array}
 \right)
\eeq
}
with
\begin{align}
\Delta_{\alpha\beta}(k,k') &= \sum_q \frac{\Delta(q)}{\sqrt{\upbeta \Omega}} A_{\alpha \beta}(\bq,\bk - \frac{\bq}{2}) \delta_{q, k-k'}\\
\tilde{\Delta}_{\alpha \beta}(k,k') &= \sum_q \frac{\Delta^*(-q)}{\sqrt{\upbeta \Omega}} A_{\beta \alpha}(-\bq,\bk - \frac{\bq}{2}) \delta_{q, k-k'}
\end{align}
We integrate out the fermions to obtain the action only in terms of the pairing field
\beq\mylabel{eqn:DeltaAction}
{\cal S}[\Delta] =  - \frac{1}{\upsilon} \sum_q \Delta^*(q) \Delta(q) - \ln\det[-G]
\eeq

We now perform a saddle point analysis of the action and look for static and homogeneous solutions via the ansatz
\beq
\Delta^{\mbox{sp}}(q) = \sqrt{\upbeta \Omega} \sqrt{2} \Delta_0 \delta_{q,0}
\eeq
where the factor of $\sqrt{2}$ is introduced for convenience. With this ansatz for the saddle point, the Green's function $G(k,k')$ is
\beq
G(k,k') = \left( 
\begin{array}{cccc}
G^p_+(k) & G^a_+(k) & 0 & 0 \\
-G^a_+(k) & G^h_+(k) & 0 & 0 \\
0 & 0 & G^p_-(k) & G^a_-(k) \\
0 & 0 & -G^a_-(k)  & G^h_-(k)
\end{array}
\right) \delta_{k,k'}
\eeq
where
\begin{align}
G^p_\alpha(k) & =  \frac{ik_n + \xi_\alpha(\bk)}{(ik_n)^2 - E_\alpha^2(\bk)} \\
G^h_\alpha(k) & =  \frac{ik_n - \xi_\alpha(\bk)}{(ik_n)^2 - E_\alpha^2(\bk)} \\
G^a_\alpha(k) & =  \frac{i \alpha \Delta_0}{(ik_n)^2 - E_\alpha^2(\bk)} 
\end{align}
with $E_\alpha(\bk) = \sqrt{\xi_\alpha(\bk)^2 + \Delta_0^2}$. The saddle point condition, after appropriate frequency sums, is
\beq\mylabel{eqn:GapEquation}
-\frac{1}{\upsilon} = \frac{1}{2 \Omega} \sum_{\bk \alpha} \frac{\tanh{\frac{\upbeta E_\alpha(\bk)}{2}}}{2 E_\alpha(\bk)}
\eeq
and agrees with the gap equation derived in ref.~\myonlinecite{Vyasanakere2011BCSBEC,Vyasanakere2011Rashbon}. The saddle point number equation is 
\beq\mylabel{eqn:NumberEquation}
\rho = \frac{1}{2 \Omega} \sum_{\bk \alpha} \left(1 - \frac{\xi_{\alpha}(\bk)}{E_\alpha(\bk)} \right)
\eeq
The values of $\Delta_0$ and $\mu$ are set by the simultaneous solution of \eqn{eqn:GapEquation} and \eqn{eqn:NumberEquation}.

Collective excitations of the system are described by fluctuations about the saddle point state. We treat them at Gaussian level by introducing ``small oscillations'' about the saddle point value of the pairing field,
\beq
\Delta(q) = \Delta^{\mbox{sp}}(q) + \eta(q)
\eeq
After some straightforward, if lengthy, algebra, the action to quadratic order in $\eta$ is
\beq
{\cal S}[\eta] = {\cal S}^{\mbox{sp}} + \half\sum_q 
\left( 
\begin{array}{cc} 
\eta^*(q) &
\eta(-q)
\end{array}
\right)
 \bPi(q)
\left( 
\begin{array}{c} 
\eta(q) \\
\eta^*(-q)
\end{array}
\right)
\eeq
where 
\begin{widetext}
{\small
\begin{equation}
\begin{split}\mylabel{eqn:PiPropagator}
\bPi(q) & = \left(
\begin{array}{cc}
\Pi_{11}(q) & \Pi_{12}(q) \\
\Pi_{21}(q) & \Pi_{22}(q) 
\end{array}
\right) \\
\Pi_{11}(q) & = \Pi_{22}(-q) = -\frac{1}{\upsilon} +  \frac{1}{\upbeta \Omega} \sum_{k,\alpha \beta} |A_{\alpha \beta}(\bq,\bk)|^2 G^p_{\alpha}(i q_\ell + ik_n, \frac{\bq}{2} + \bk) G^h_{\beta}( ik_n, -\frac{\bq}{2} + \bk) \\
\Pi_{12}(q) &= \Pi_{21}(q) = - \frac{1}{\upbeta \Omega} \sum_{k,\alpha \beta} \alpha \beta  |A_{\alpha \beta}(\bq,\bk)|^2 G^a_{\alpha}(i q_\ell + ik_n, \frac{\bq}{2} + \bk) G^a_{\beta}( ik_n, -\frac{\bq}{2} + \bk) = \Pi_{12}(-q) = \Pi_{21}(-q)
\end{split}
\end{equation}
}
\end{widetext}

Collective excitations of a superfluid can be conveniently described in terms of spatio-temporally dependent  phase and amplitude oscillations.  We, therefore, express $\eta$ in terms of two other real fields $\zeta$ (amplitude fluctuation) and $\phi$ (phase fluctuation) as
\beq
\eta(q) = \Delta_0 \left( \zeta(q) + i \phi(q) \right) 
\eeq
with $\zeta(-q) = \zeta^*(q)$ and $\phi(-q) = \phi^*(q)$. The action in terms of these two fields is 
\beq
{\cal S}[\zeta,\phi] = {\cal S}^{sp} + \half\sum_q \left(\begin{array}{cc}\zeta^*(q) & \phi^*(q) \end{array} \right) \bGam(q) \left( \begin{array}{cc} \zeta(q) \\ \phi(q) \end{array} \right)
\eeq
where, using \eqn{eqn:PiPropagator}, we find
\begin{align}
\bGam(q) &= \left( 
\begin{array}{cc}
\Gamma_{\zeta \zeta}(q) & \Gamma_{\zeta \phi}(q) \\
\Gamma_{\phi \zeta}(q) & \Gamma_{\phi \phi}(q)
\end{array}
\right) \\
\Gamma_{\zeta \zeta}(q) &= \Delta_0^2 \left(\Pi_{11}(q) + \Pi_{11}(-q) + 2 \Pi_{12}(q) \right) \\
\Gamma_{\zeta \phi}(q) &=   i \Delta_0^2 \left(\Pi_{11}(q) - \Pi_{11}(-q) \right) = -\Gamma_{\phi \zeta}(q) \\
\Gamma_{\phi \phi}(q) & = \Delta_0^2  \left(\Pi_{11}(q) + \Pi_{11}(-q) - 2 \Pi_{12}(q) \right) 
\end{align}
We now preform the necessary frequency sums to obtain expressions for the $\Gamma$s. Here and henceforth in this paper, we focus at zero temperature ($T=0$) and ``small'' $q$, and do not show the lengthy expressions valid for any temperature and $q$. For small $q$ at $T=0$, we have, 
\begin{align} \mylabel{eqn:GammaDefs}
\Gamma_{\phi \phi}(iq_\ell, \bq) &=  q_i K^s_{ij} q_j - Z (i q_\ell)^2 \\
\Gamma_{\zeta \phi}(iq_\ell, \bq) &= - i q_\ell X   \\
\Gamma_{\zeta \zeta}(iq_\ell, \bq) &= U + q_i V_{ij} q_j - W (i q_\ell)^2 
\end{align}
where the quantities $K^s,Z,X, U, V, W$ depend on the saddle point values of $\Delta_0$ and $\mu$. $K^s_{ij}$ is the phase stiffness given by
\begin{equation}\mylabel{eqn:PhaseStiffness}
\begin{split}
K^s_{ij} & = \frac{\Delta_0^2}{2 \Omega} \sum_{\bk \alpha} \frac{v^\alpha_i(\bk) v^\alpha_j(\bk)}{4 E_\alpha^3(\bk)} \\
& + \frac{2 \Delta_0^2}{\Omega} \sum_{\bk} \frac{\left(\varepsilon_+(\bk) - \varepsilon_-(\bk)\right)^2}{2 E_+(\bk) E_-(\bk) \left(E_+(\bk) + E_-(\bk) \right)} S_{ij}(\bk)
\end{split}
\end{equation}
where $v^\alpha_i(\bk) = \frac{\dou \varepsilon_\alpha(\bk)}{\dou k_i}$, and $S_{ij}(\bk)$ is a tensor that defines the singlet amplitude $A_{+-}(\bq,\bk)$ for small $\bq$ as
\beq
|A_{+-}(\bq,\bk)|^2 = |A_{-+}(\bq,\bk)|^2 \approx q_i S_{ij}(\bk) q_j.
\eeq
It must be noted that extensive use of the properties of $A_{\alpha \beta}(\bq,\bk)$ is made in arriving at this expression for the phase stiffness tensor that is valid for {\em any} gauge field. The other quantities in \eqn{eqn:GammaDefs},
\begin{equation}\mylabel{eqn:UVWXYZdefs}
\begin{split}
Z & = \frac{\Delta_0^2}{2 \Omega} \sum_{\bk \alpha} \frac{1}{4 E_\alpha^3(\bk)} \\
X &=  \frac{\Delta_0^2}{2 \Omega} \sum_{\bk \alpha} \frac{\xi_\alpha(\bk)}{2 E_\alpha^3(\bk)}\\
U &= \frac{\Delta_0^4}{2 \Omega} \sum_{\bk \alpha} \frac{1}{E_\alpha^3(\bk)} \\
W &= Z -  \frac{\Delta_0^4}{2 \Omega} \sum_{\bk \alpha} \frac{1}{4 E_\alpha^5(\bk)}. \\
\end{split}
\end{equation}
We have not shown the expression for $V_{ij}$ since it will not be used in the discussion below.

The dispersion of the excitations can be obtained by first analytically continuing $i q_\ell \rightarrow \omega^+$ to real frequencies and solving $\det{\bGam(\omega^+,\bq)}=0$. We obtain two modes for a given $\bq = q \hat{\bq}$, one is a {\em gapless} sound mode and other is the {\em gapped} Anderson-Higgs mode. The speed of sound along direction $\hat{\bq}$ is given by
\beq\label{eqn:Cs}
c^2_s(\hat{\bq}) = \frac{\hat{q}_i K^s_{ij} \hat{q}_j}{Z + \frac{X^2}{U}} 
\eeq
and the mass of the Anderson-Higgs mode $M_{AH}$ is obtained as
\beq\label{eqn:MAH}
M_{AH}^2 = \frac{ZU + X^2}{Z W}
\eeq
It must be noted that the amplitude and phase modes are coupled\mycite{Engelbrecht1997}; their coupling is determined by the quantity $X$.

Equations \ref{eqn:PhaseStiffness}, \ref{eqn:Cs} and \ref{eqn:MAH} are
the key results of this paper for the collective excitations of
spin-orbit coupled superfluids that are applicable to {\em any} Rashba
gauge field and scattering length at zero temperature. We have not
shown the finite temperature results here to avoid lengthy
expressions. In the remainder of the paper, we illustrate the physics
of these formulae using the spherical gauge field (next section) and
explore the consequences of our results particularly for the
rashbon-BEC (sec.~\ref{RBECdiscussion}).

\section{Collective excitations for the spherical gauge field}
\mylabel{SphericalResults}

In this section we discuss collective excitations of superfluids
realized in a spherical gauge field with $\blam = \lambda
\left(\frac{1}{\sqrt{3}},\frac{1}{\sqrt{3}}, \frac{1}{\sqrt{3}}
\right)$ as noted earlier. The two body problem  for this
gauge field was exhaustively investigated in
ref.~\myonlinecite{Vyasanakere2011TwoBody} where an analytical expression for the
binding energy valid for {\em any scattering length} is derived along
with an analytical expression for the bound state wave function. The
binding energy of the rashbon\mycite{Vyasanakere2011TwoBody} is
\beq\mylabel{eqn:ERspherical}
E^R = \frac{\lambda^2}{3}
\eeq
and the rashbon mass (in units of fermion mass) is\mycite{Vyasanakere2011Rashbon}
\beq\mylabel{eqn:MRspherical}
m^R = \frac{3}{7}(4 + \sqrt{2})
\eeq
A route to experimental realization of this gauge field has recently been  suggested.\mycite{Anderson2011} A detailed study of two-body scattering from a finite range box potential is carried out in ref.~\myonlinecite{Cui2011}.

\subsection{Analytical Results}
Analytical results can be obtained in two regimes of $\lambda$. These correspond to $\lambda \ll \kf$, and the other to $\lambda \gg \max{(\kf, 1/\as)}$.

\subsubsection{$\lambda \ll \kf$}
Two regimes of $\as$ are tractable analytically for this regime of $\lambda$, both of which are well known; we state them here for the sake of completion.

\noindent
I.~$\as < 0, |\kf \as| \ll 1$: This regime is studied in detail in ref.~\myonlinecite{Vyasanakere2011BCSBEC}. The chemical potential in this regime is set by the value of the noninteracting system (which falls by an amount proportional to $\frac{\lambda^2}{\kf^2}$). The gap $\Delta_0$ is essentially unaltered from the well known BCS value. Under these conditions, we obtain the phase stiffness to be $\frac{\rho}{4}$ with a fall of order $\frac{\lambda^2}{\kf^2}$. The leading term in the speed of sound is $\kf/\sqrt{3}$ as shown by Anderson\mycite{Anderson1958} (with a fall proportional to $\lambda^2/\kf^2$) and the Anderson-Higgs mass is exponentially small. This limit corresponds essentially to the BCS limit studied in ref.~\myonlinecite{Engelbrecht1997}.

II.~$\as>0, \kf \as \ll 1$: This corresponds to the usual BEC regime (ref.~\myonlinecite{Engelbrecht1997,Pethick2004}). Here the chemical potential  $\mu = -\frac{1}{2 \as^2} + 2 \pi \as \rho$ and the gap $\Delta_0^2 = \frac{4 \pi \rho}{\as} $. The phase stiffness $K^s = \frac{\rho}{4}$, speed of sound is $c_s^2 = 2 \pi \rho \as$, the $M_{AH} = \frac{4}{\as^2}$. In this regime, the amplitude and the phase modes are strongly mixed.

\subsubsection{$\lambda \gg \kf$ and $\lambda \gg \frac{1}{\as}$}

This is the regime of interest and corresponds to the rashbon BEC. In this regime, we report new results for the gap
\beq\mylabel{eqn:SphericalDelta}
\Delta_0^2 = \frac{2 \pi}{ \rho} \frac{\lambda}{\sqrt{3}}
\eeq
and the chemical potential 
\beq\mylabel{eqn:SphericalMu}
\mu = - \frac{E^R}{2} + \pi \rho \frac{\sqrt{3}}{\lambda}.
\eeq

By an analysis of the expression for the phase stiffness (\eqn{eqn:PhaseStiffness}) which is isotropic for this gauge field, we find that 
\beq\mylabel{eqn:SphericalKs}
K^s = \frac{\rho}{2 m^R}
\eeq
{\em precisely as conjectured in the introductory section} (see below for further discussion).
Additional analysis provides
\beq\mylabel{eqn:Sphericalcs}
c_s^2 = \frac{2 \pi \rho}{m^R} \left(\frac{\sqrt{3}}{\lambda} \right)
\eeq
and 
\beq\mylabel{eqn:SphericalMAH}
M_{AH} = \frac{2}{3} \lambda^2.
\eeq
 As expected, the leading terms for all the quantities of interest are {\em independent of the scattering length between the fermions}; scattering length corrections (which we do not show) appear as powers of $(1/\lambda \as)$, which in this regime are small. We emphasize that in this RBEC regime the amplitude and the phase mode are strongly coupled, just like in the usual BEC regime.

\subsection{Numerical Results}

In this section we show the results of numerical calculations of evolution of $K^s$, $c_s$ and $M_{AH}$  with increasing $\lambda$ for several scattering lengths. 

\begin{figure}
\centerline{\includegraphics[width=\myfigwidth]{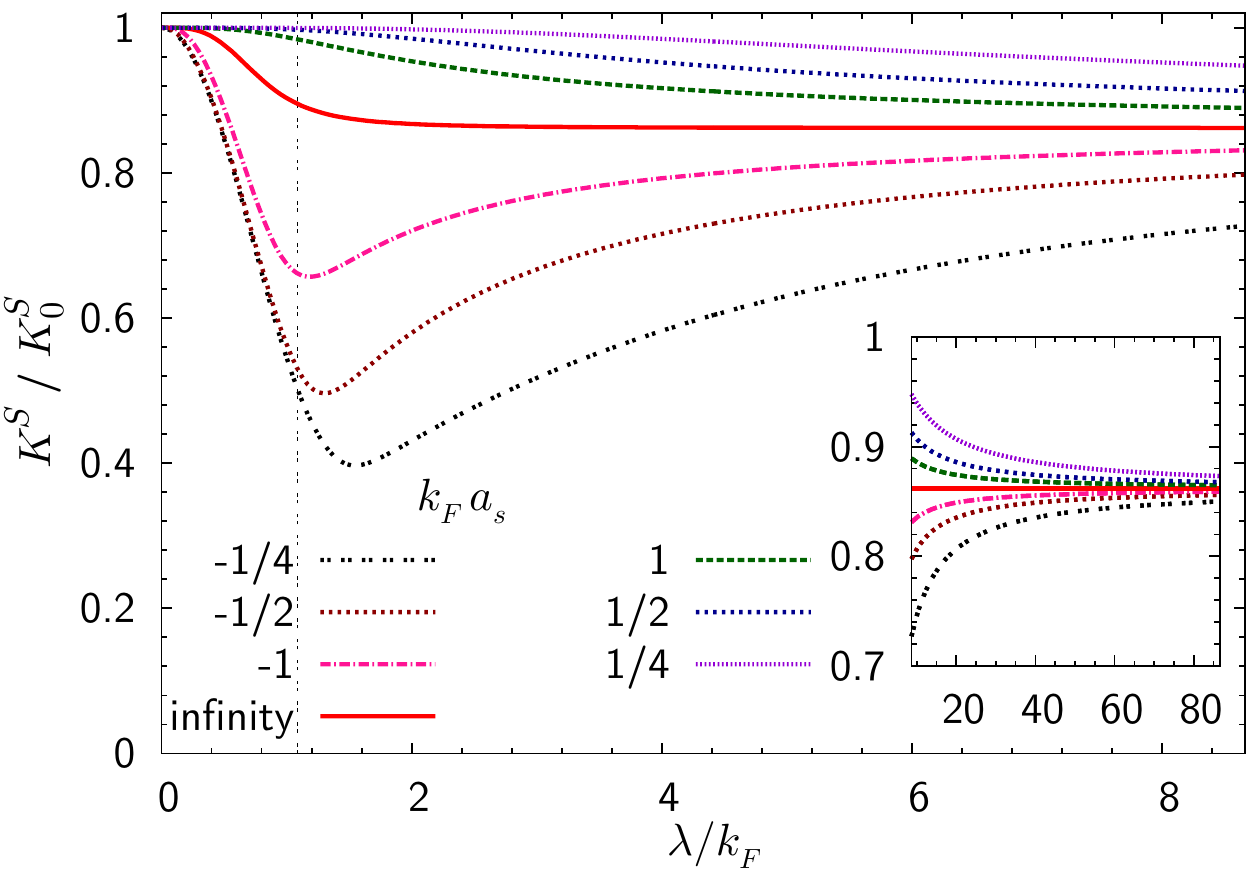}}
\caption{(color online) {\bf Phase stiffness} - Evolution of phase stiffness $K^s$ with increasing $\lambda$ for the spherical gauge field for various scattering lengths. $K^s_0= \rho/4$. The inset shows that $K^s/K_0$ tends to $2/m^R$ for large $\lambda$ demonstrating the emergent Galilean invariance. The dashed vertical line corresponds to $\lambda=\lambda_T$ where there is a change in the topology of the Fermi surface of the non-interacting system.\mycite{Vyasanakere2011BCSBEC}}
\mylabel{fig:PhaseStiffness}
\end{figure}

\subsubsection{Superfluid Phase Stiffness}

\Fig{fig:PhaseStiffness} shows a plot of the phase stiffness as a function of $\lambda$ for various scattering lengths. We see that for small negative scattering lengths, the behaviour of $K^s$ is non-monotonic; it decreases with increasing $\lambda$ and attains a minimum near $\lambda \gtrsim \lambda_T$. This is fully consistent with the finding of ref.~\myonlinecite{Zhou2011} for the EO gauge field. The new aspect uncovered in our work is that for $\lambda \gg \max{(\kf,1/\as)}$, the phase stiffness tends to that of a {\em collection of interacting rashbons} in exactly same way as the motivating conjecture of this paper. In other words, $K^s(\lambda \rightarrow \infty) = \frac{\rho_R}{m^R}$ where $\rho_R = \rho/2$ is the rashbon number density. The physics behind this is that the rashbon dispersion $\varepsilon_R(\bq) = -E^R + \frac{q^2}{2 m^R}$ is Galilean invariant, and hence the phase stiffness as found at $\lambda \rightarrow \infty$ is consistent with Leggett's result\mycite{Leggett1998,Leggett2006}. This is a remarkable feature, and corresponds to an {\em emergent infrared symmetry}, i.~e., {\em in the presence of interactions however small, the system organizes itself to posses a larger symmetry at low energies!} A important point that can be inferred is that the nonzero phase stiffness implies that rashbons are {\em interacting} bosons. The nature of the interaction is uncovered in the next section.

\begin{figure}
\centerline{\includegraphics[width=\myfigwidth]{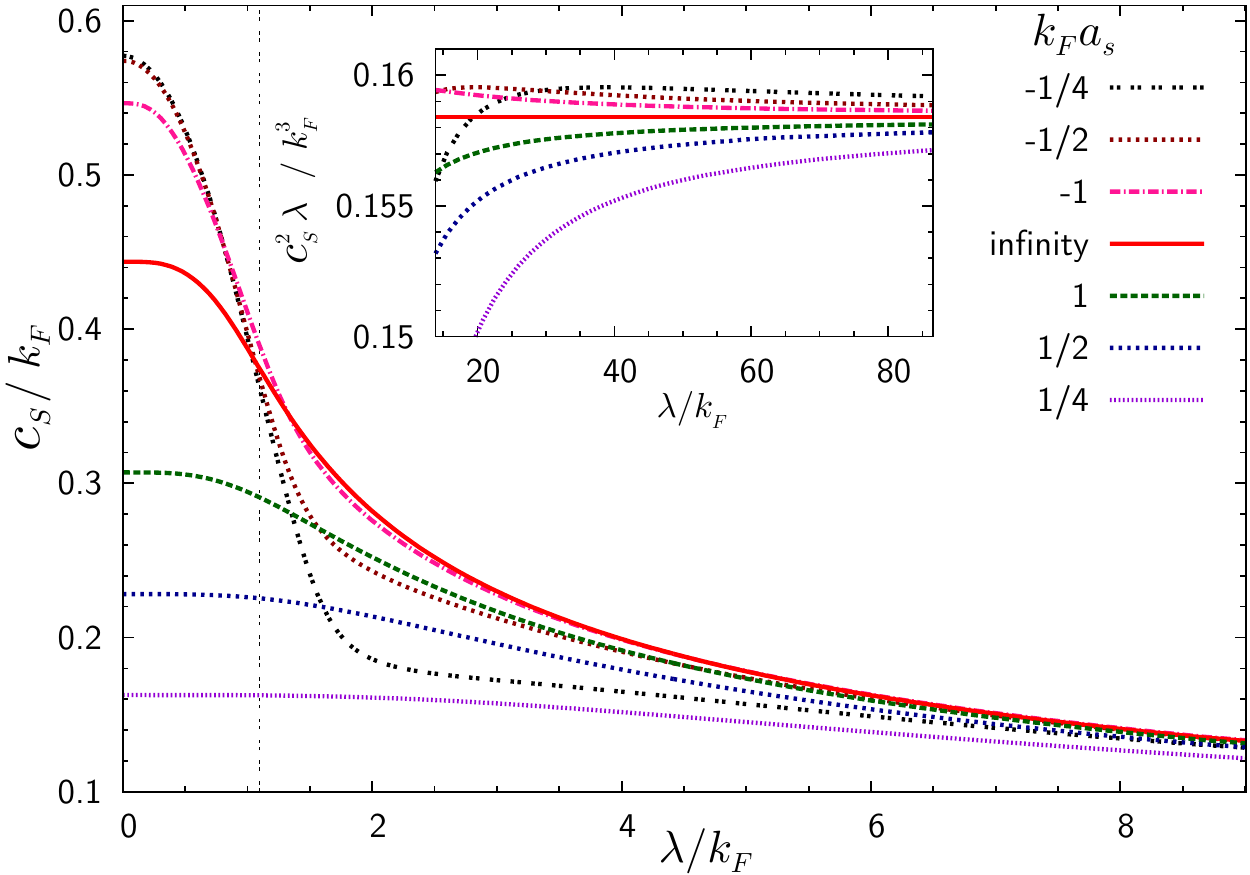}}
\caption{(color online) {\bf Sound speed} - Evolution of the sound speed $c_s$ with increasing $\lambda$ for the spherical gauge field for various scattering lengths. The inset shows that $c_s^2$ has the behaviour obtained in \eqn{eqn:Sphericalcs}, independent of the scattering length. The dashed vertical line corresponds to $\lambda=\lambda_T$ where there is a change in the topology of the Fermi surface of the non-interacting system. }
\mylabel{fig:SoundSpeed}
\end{figure}

\subsubsection{Sound Speed}

The variation of the sound speed with increasing $\lambda$ is shown in \fig{fig:SoundSpeed}. We see that there is a monotonic decrease in the sound speed with increasing $\lambda$ for all scattering lengths. At large $\lambda$, the sound speed is inversely proportional to $\lambda$ as obtained analytically (see \eqn{eqn:Sphericalcs}). Again, that there is sound propagation in the medium suggests the presence of interactions between the rashbons.

\subsubsection{Mass of the Anderson-Higgs boson}

For small gauge coupling ($\lambda \ll \kf$ ) $M_{AH}$ corresponds to the gap of the amplitude mode for small negative scattering lengths. This mass grows with increasing $\lambda$ albeit with some features near $\lambda \sim \lambda_T$ for small negative scattering lengths. At large $\lambda$ we find the expected $\lambda^2$ behaviour.

The key result of this section is that at large $\lambda$, the system behaves like a Galilean invariant collection of interacting rashbons. Since this regime is the raison d'etre of this paper, we do not pause to consider the interesting regime of $\lambda \sim \lambda_T$ which no doubt contains rich physics.

\begin{figure}
\centerline{\includegraphics[width=\myfigwidth]{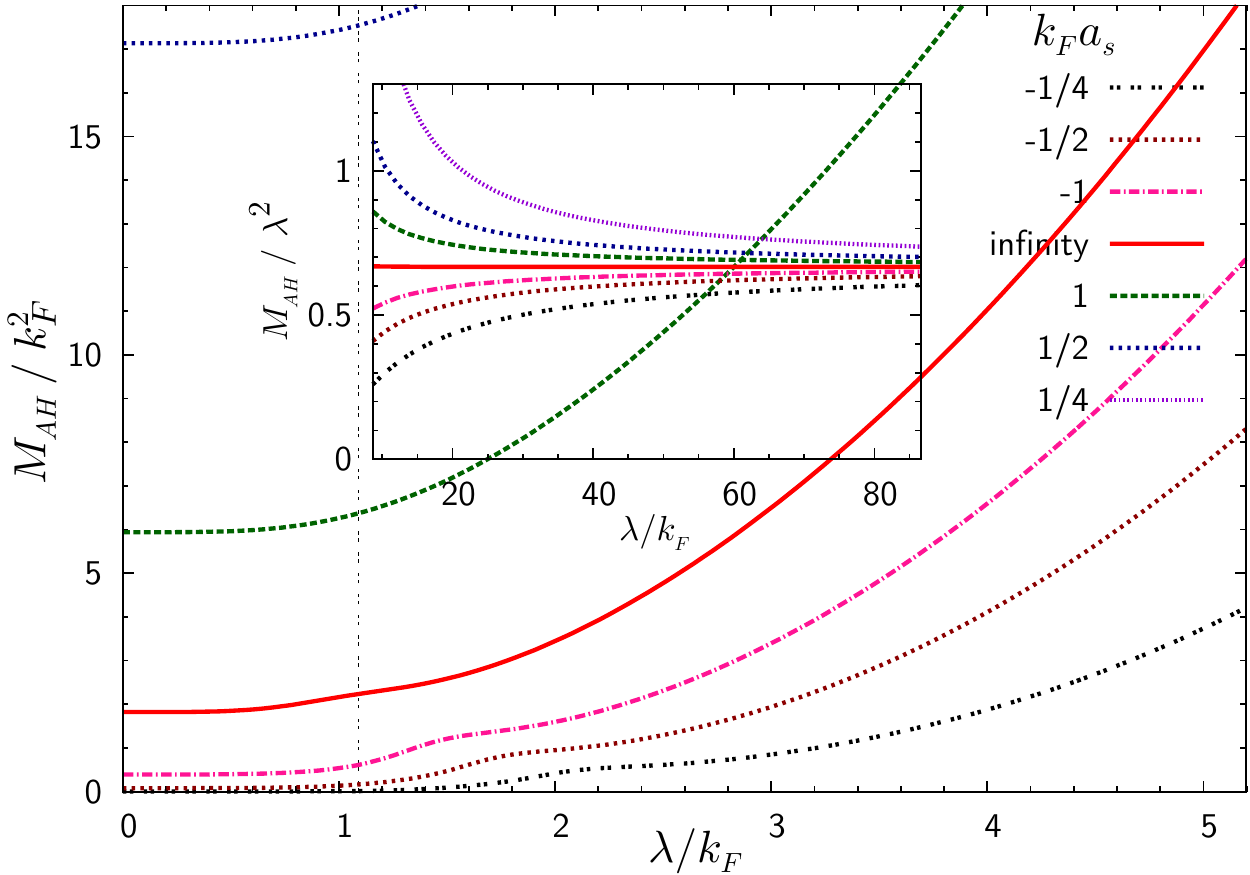}}
\caption{(color online) {\bf Mass of the Anderson-Higgs boson} - Evolution of mass of the Anderson-Higgs boson $M_{AH}$ with increasing $\lambda$ for the spherical gauge field for various scattering lengths is shown in \Fig{fig:HiggsMass}. The inset shows that $M_{AH}$ goes as $\lambda^2$, independent of the scattering length (\eqn{eqn:SphericalMAH}). The dashed vertical line corresponds to  $\lambda=\lambda_T$ where there is a change in the topology of the Fermi surface of the non-interacting system. }
\mylabel{fig:HiggsMass}
\end{figure}

\section{Properties of Rashbon Bose-Einstien Condensates (RBEC)}
\mylabel{RBECdiscussion}

That the system evolves to a collection of interacting rashbons with increasing $\lambda$ is conclusively demonstrated in the previous section. The rashbon dispersion derived in ref.~\myonlinecite{Vyasanakere2011Rashbon} provides the kinetic energy of the rashbons. What about their interactions? Interestingly, the results of the previous section allow us to answer this question.

Recall from the Bogoliubov theory\mycite{Abrikosov1965} that a collection of bosons of  mass $m_B$ with number density $\rho_B$ and a contact interaction described by a scattering length $a_B$ has a superfluid ground state at zero temperature. The chemical potential of this system is
\beq\mylabel{eqn:BosonMu}
\mu_B = \frac{4 \pi a_B}{m_B} \rho_B
\eeq
and the speed of sound is
\beq\mylabel{eqn:Bosoncs}
c_s^B = \sqrt{ \frac{\mu_B}{m_B}} = \sqrt{\frac{4 \pi a_B \rho_B}{m_B^2}} .
\eeq

From \eqn{eqn:SphericalMu}, the rashbon chemical potential $\mu^R$ (measured from the bottom of the rashbon band at $-E^R$) is
\beq\mylabel{eqn:RashbonMu}
\mu^R = 2 \pi \rho \frac{\sqrt{3}}{\lambda}
\eeq
We see immediately that the speed of sound obtained in \eqn{eqn:Sphericalcs} is {\em consistent with \eqn{eqn:Bosoncs} from  Bogoliubov theory} 
\beq\mylabel{eqn:Rashboncs}
c_s^2 = \frac{\mu^R}{m^R}
\eeq
This clearly demonstrates that the rashbon BEC is a condensate of rashbons interacting with a contact interaction. Writing
\beq
c_s^2 = \sqrt{\frac{4 \pi a_R \rho_R}{(m^R)^2}}
\eeq
allows us to calculate the rashbon-rashbon scattering length as
\beq
a_R = \frac{3 \sqrt{3} (4 + \sqrt{2})}{7} \frac{1}{\lambda} 
\eeq
which is approximately equal to $\frac{4}{\lambda}$. This result is remarkable in the following sense that the effective interaction between rashbons is determined by a scale $\lambda$ that enters the kinetic energy of the constituent fermions, and {\em not} by the interaction between the constituent fermions (scattering length $\as$)! 

We emphasize that although our arguments used the spherical gauge fields, the results obtained are applicable to other gauge field configurations described by a general vector $\blam = \lambda \hat{\blam}$ (except the extreme prolate gauge field which has only one nonvanishing component, see ref.~\myonlinecite{Vyasanakere2011BCSBEC}). For a generic gauge field, the rashbon chemical potential will be
\beq
\mu^R = M(\hat{\blam}) \frac{\rho}{\lambda}
\eeq
where $M(\hat{\blam})$ is a dimensionless number that depends on $\hat{\blam}$, and the anisotropic speed of sound in the $i$-direction will be
\beq
c_s^2(i) = \frac{\mu^R}{m_i^R}
\eeq
where $m_i^R$ is the anisotropic rashbon mass\mycite{Vyasanakere2011Rashbon} that depends, again, on $\hat{\blam}$. The rashbon-rashbon scattering length will be 
\beq
a_R = \frac{N(\hat{\blam})}{\lambda}
\eeq
where $N(\hat{\blam})$ is dimensionless number determined by $\hat{\blam}$
The low energy properties of the rashbon BEC are similar to those of the usual Bogoliubov Bose fluid; in fact, generically, RBEC is a superfluid of anisotropically dispersing rashbons interacting with a contact potential described by a scattering that depends inversely on the spin orbit coupling strength of the fermions. It must be noted that accurate determination of $N(\hat{\blam})$ may require further self consistent treatment of the theory.\mycite{Hu2006,Diener2008}

\begin{figure}
\centerline{\includegraphics[width=\myfigwidth]{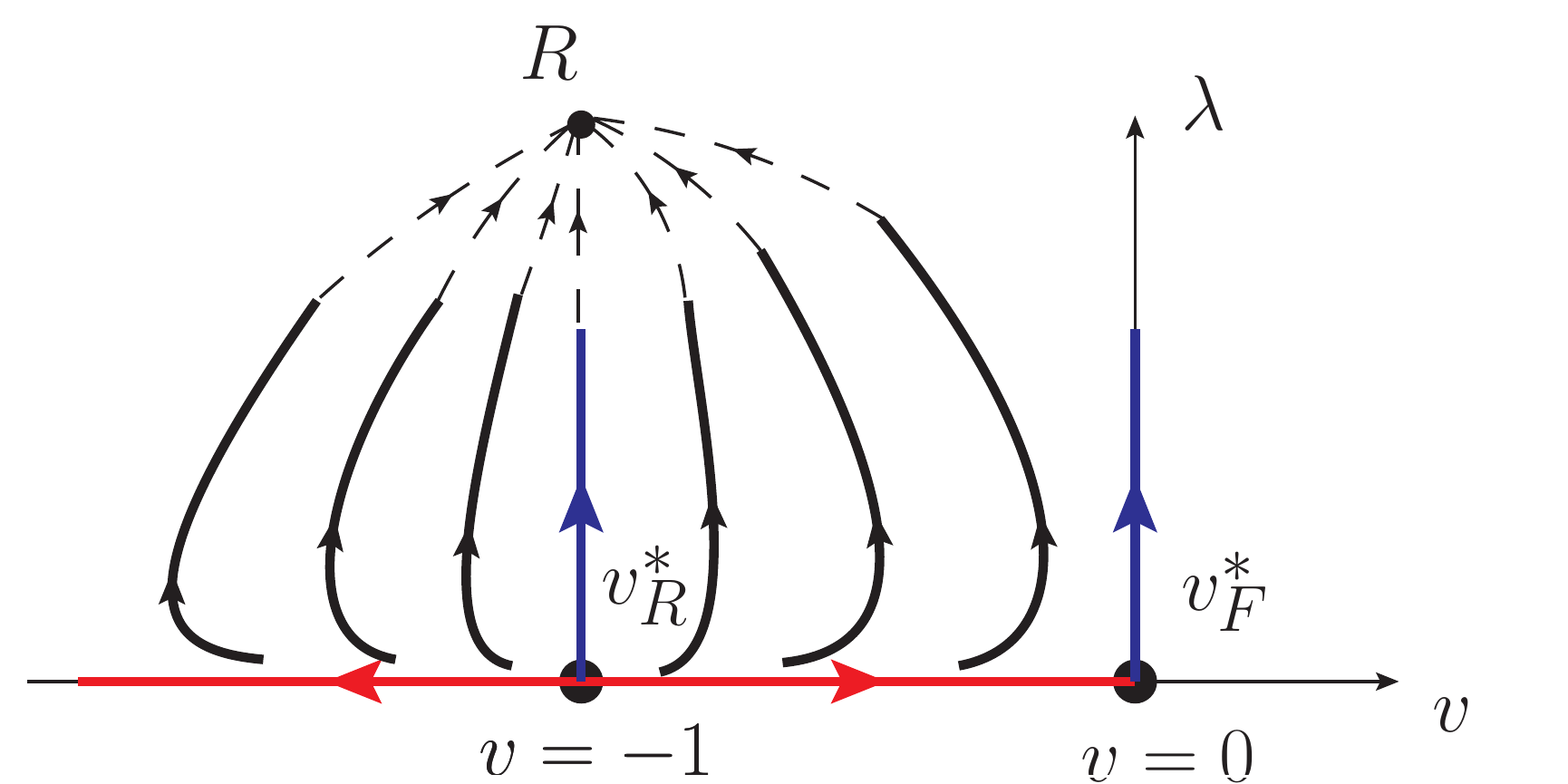}}
\caption{(color online) {\bf Schematic two-body  RG flow diagram} - The rashbon state corresponds to the stable fixed point $R$ at $\lambda = \infty$ and $v = -1$. Flow from any point with $\lambda \ne 0$ and $v \ne 0$ reaches $R$.}
\mylabel{fig:RGflow}
\end{figure}

\section{Summary}
\mylabel{Summary}

In this paper, we explore the properties of the superfluids induced by non-Abelian gauge fields focusing on their collective excitations. We present results for superfluid phase stiffness, sound speed and Anderson-Higgs mass valid for {\em any} Rashba gauge field and scattering length. Our main results are
\begin{itemize}
\item Superfluid phase stiffness has non-monotonic behaviour with increasing $\lambda$, the scale of the spin-orbit interaction. This is in agreement with an earlier report\mycite{Zhou2011} of superfluid density for the EO gauge field.
\item A new result is that for large gauge coupling, i~e., in the rashbon BEC, the superfluid phase stiffness is determined by the rashbon mass\mycite{Vyasanakere2011Rashbon}. This arises from {\em an emergent  Galilean invariance} at  infrared energies for large gauge couplings, and the phase stiffness  is consistent with Leggett's result.
\item The sound speed decreases monotonically with increasing gauge coupling. At large gauge coupling it goes as $\lambda^{-1/2}$. The Anderson-Higgs mass increases with increasing $\lambda$ and goes as $\lambda^2$ in the rashbon-BEC.
\item A key outcome of this work is that we show that the rashbon-BEC can be described as a collection of anisotropically dispersing rashbons interacting via a contact interaction. We obtain  an analytical expression for the rashbon-rashbon interaction for the spherical gauge field showing that it goes as $\lambda^{-1}$. We argue that this result is true for a generic gauge field (spin-orbit interaction).
\end{itemize}

We conclude the paper by revisiting the RG flow diagram of the two body problem introduced in ref.~\myonlinecite{Vyasanakere2011TwoBody}. \Fig{fig:RGflow} is a schematic RG flow diagram in the $\lambda$-$\upsilon$ plane for the two-particle problem. The key point is that flow from any point with $\lambda \ne 0$ {\em and} $\upsilon \ne 0$ reaches $R$ which is the stable rashbon fixed point corresponding to $\lambda = \infty$ and $\upsilon = -1$. Indeed, the properties of the state attained by a finite density of fermions at large $\lambda$ is controlled by the rashbon fixed point; it is therefore a weakly interacting gas of rashbons -- the rashbon BEC.

\subsection*{Acknowledgement}

JV acknowledges support from CSIR, India via a JRF grant.  VBS is
grateful to DST, India (Ramanujan grant), DAE, India (SRC grant) and IUSSTF
for generous support.

\bibliography{refCollective_nagf}

\end{document}